\documentclass[aps,prl,twocolumn,superscriptaddress]{revtex4}

\usepackage{graphicx}
\usepackage{latexsym}
\usepackage{amsfonts}

\newcommand{\ket}[1]{| #1 \rangle}

\begin{document}

\title{Quantum interference interpreted classically through application of Berry's phase}

\author{M. J. Rave}
\affiliation{Department of Chemistry and Physics, Western Carolina University,
Cullowhee, North Carolina 28723}

%\author{M. J. Rave}

%\institute{M. J. Rave \at
              %Department of Chemistry and Physics, Western Carolina University, Cullowhee, NC 28723 \\
              %\email{mrave@email.wcu.edu}}

\date{\today}

%\maketitle

\begin{abstract}
We show that quantum interference can be classically interpreted in terms of a phase invariant quantity, not unlike the Berry's phase.  Under this interpretation, closed loops in time become fundamental quantum entities, and all quantum states become periodic.  Decoherence is then seen to occur naturally as a consequence.  This formalism, although counterintuitive, provides a useful way of assigning ``classical'' meaning to quantum probabilities.
%\keywords{Berry's phase \and quantum interference \and quantum probability \and decoherence \and closed timelike curves}
%\PACS{02.50.Cw \and 03.65.Ud \and 03.65.Vf \and 03.65.Yz}
\end{abstract}

\maketitle

\section{Introduction}
\label{sec:Intro}
To many, quantum interference (QI) is the most mysterious aspect of quantum mechanics; in Feynman's words, it ``contains the \emph{only} mystery'' \cite{Feynman}.  The implication is that all of the non-intuitive aspects of quantum mechanics (such as wave-particle duality, non-commuting observables, the EPR ``paradox'' \cite{Einstein}, or Bell's inequality \cite{Bell}) are manifestations of a single non-classical phenomenon, namely QI.  It has become common to follow Feynman's lead and say that there is no way to \emph{explain} QI
in a ``classical way'' \cite{Feynman}; by this it is meant that QI involves probabilities (cross-terms) that seemingly do not appear in classical probability theory.  However, recent work \cite{Kirkpatrick} has shown that certain decidedly classical systems can exhibit superficially non-classical probabilities and pseudo-QI effects.  Thus it becomes natural to reexamine QI in an attempt to assign a classical meaning to non-classical terms.  Clearly, an understanding of \emph{why} such non-classical probabilities show up in quantum systems is important in developing any meaningful interpretation of quantum mechanics itself.  In the words of Mermin, ``an acceptable notion of probabilities as objective properties of individual systems'' may ``sweep all the puzzles of quantum mechanics'' under the rug \cite{Mermin}.

Part of the perceived difficulty in interpreting quantum mechanics lies in the structure of quantum mechanics: what concepts, if any, should be considered fundamental?  For example, are state vectors more fundamental than operators?  At first the question seems moot, since the Schr\"{o}dinger and Heisenberg pictures are mathematically equivalent (up to a change of basis).  But consider that all observables correspond to Hermitian operators: does this mean that such operators have a special place in the hierarchy of quantum ideas?

The concept of the Berry's phase (BP) \cite{Berry} has shed new light onto such philosophical questions.  In 1984 Berry showed that although the phase of a given state vector (ket) cannot have physical meaning, a phase invariant quantity (now called the Berry's phase) can always be constructed in the Hilbert space of the state vectors, and this phase is measurable.  Berry's idea came as a revelation in the mid-1980's, for the standard dogma of the time was that for a quantity to be measurable, it must be the eigenvalue of some Hermitian operator.  Berry showed that Berry's phase is measurable, but occurs as a result of the topology of the Hilbert space in which the BP resides.  Hence there are no operators that can be associated with the BP.  In Resta's words, Berry's phase is an ``exotic'' observable \cite{Resta}.  Being observable, then, is there any reason to place the BP on par with state vectors and/or Hermitian operators as quantum mechanical constructs?

\section{Berry's phase}
\label{sec:Berry}
Berry's phase is usually introduced within the context of a quantum system governed by a parametric Hamiltonian $\mathcal{H}(x)$ \cite{Resta}.  (This idea is convenient for didactic purposes, but is not required for the BP to manifest itself.)  For now we will define the BP as a phase shift in a system's state vector $\ket{\Psi(x)}$ that appears after the system has traced a closed loop in the space of some parameter $x$.  Consider a system that evolves cyclically as $\ket{\Psi(x_1)}\rightarrow\ket{\Psi(x_2)}\rightarrow\ket{\Psi(x_3)}\rightarrow\ket{\Psi(x_1)}$.  (We will refer to this ordered sequence of states as a ``path'' even though it involves a discrete number of states.  The path is considered closed because the final state of the system is the same as the initial state.)  For this path the BP ($\gamma$) is defined to be \cite{Resta}
\begin{eqnarray} \label{BP}
\gamma = &&-\textrm{Im} \ln [ \langle \Psi(x_1) \ket{\Psi(x_3)} \nonumber \\
 && \langle \Psi(x_3) \ket{\Psi(x_2)}\langle \Psi(x_2) \ket{\Psi(x_1)} ].
\end{eqnarray}
The most important feature of this definition is that $\gamma$ is phase invariant (sometimes called ``gauge invariant'').  By this we mean that individual phase choices for particular kets don't matter; the BP still comes out the same for a given path.  This can be seen by making the ``gauge transformations'' $\ket{\Psi(x_1)}\rightarrow e^{i \theta_1} \ket{\Psi(x_1)}, \ket{\Psi(x_2)}\rightarrow e^{i \theta_2} \ket{\Psi(x_2)}, \ket{\Psi(x_3)}\rightarrow e^{i \theta_3} \ket{\Psi(x_3)}$ and recalculating $\gamma$.

Because it is invariant to phase choice, the discrete BP (and its continuum limit counterpart) is a measurable quantity.  For example, in systems of Bloch electrons the presence of the BP is seen as either an electron self-interference effect (\cite{Loss},\cite{Geller}) or as a half-integer quantum Hall effect\cite{Zhang}.  The BP appears in these \emph{particular} systems because of the torus topology of the Brillouin zone; $k$'s (Bloch pseudomomenta) that differ by a reciprocal lattice vector are considered equal.  Similarly, in all other situations in which a BP appears (electric polarization \cite{Kingsmith}, spin-wave dynamics \cite{Niu}, or even molecular machines \cite{Astumian}) the state vectors traverse a closed loop in the space of some parameter.  The importance of such closed loops motivates further research into placing the BP into the structure of quantum theory.

\section{Quantum probabilities}
\label{sec:Quantum}
Recent work \cite{Anastopoulos} has shown that there is a link between the BP and quantum probability.  Specifically, in employing a consistent histories \cite{Griffiths} formulation of quantum mechanics, one finds that the BP becomes ``the main building block of the decoherence functional'' \cite{Anastopoulos}, which itself contains complete information about a quantum system.  Put another way, for each quantum mechanical history there can be associated (mapped) a unique BP.  The authors of \cite{Anastopoulos} show that ``all physically relevant amplitudes'' are based on the BP, and conclude that the presence of complex phases leads to non-classical probability terms (i.e. QI).  We do not disagree with this result but suggest that there $is$ a way to interpret QI \emph{classically}, if one is willing to accept a paradigm shift in analyzing quantum transitions.

For the purposes of our discussion, we will define QI through an examination of probability.  We say that a system involves \emph{classical} probability if the system always obeys the addition law \cite{Rozanov}
\begin{equation} \label{classical law}
P(A_1 \cup A_2)=P(A_1)+P(A_2),
\end{equation}
with $A_1$ and $A_2$ being mutually exclusive events, and $P(A_1 \cup A_2)$ being the union of these events.  In contrast, \emph{quantum} systems are immediately identifiable by the presence of equalities such as
\begin{equation} \label{interference}
P(A_1 \cup A_2) = P(A_1)+P(A_2)+f(P(A_1),P(A_2)),
\end{equation}
where $f(P(A_1),P(A_2))$ is, in general, non-zero. In analogy with the mathematics of interfering waves, we call terms like $f(P(A_1),P(A_2))$ ``interference terms'' \cite{Feynman}.

Note that this definition of classical probability is an operational one; we do not enter into any debate between frequentists and Bayesians \cite{Bland} as to the epistemological meaning of probability.   Nor do we refer to past and/or future events in order to give meaning to probabilities, as some have done (e.g. \cite{Weizsacker}).  Our goal is simply to show that quantum interference terms become classically explicable (and hence cease to be ``interference'' terms at all) if one looks at ``closed loops in time'' as fundamental quantum entities.

\section{Product of amplitudes}
\label{sec:Product}
We represent state vectors by kets in a complex vector space.  In such a space $\ket{\Psi}$ and $e^{i \theta}\ket{\Psi}$ are said to correspond to the same physical state, since they lead to identical expectation values for identical Hermitian operators.  Now consider two distinct states $\ket{\Psi_1}$ and $\ket{\Psi_2}$.  The inner product $\phi_{21} \equiv \langle \Psi_2 \ket{\Psi_1}$ is called a probability amplitude, since the probability of the transition $\Psi_1\ \rightarrow \Psi_2$ is taken to be $|\langle \Psi_2 \ket{\Psi_1}|^2$.  Since $\phi_{21} \in \mathbb{C}$, we can write the amplitude in modulus/phase form as
\begin{equation}
\phi_{21} = r_{21} e^{-i \theta_{21}}.
\end{equation}
The phase difference $\theta_{21}=-\textrm{Im} \ln \langle \Psi_2 \ket{\Psi_1}$ is not a ``Berry's phase'' since we do not yet have a closed loop.

Now assume that we have \emph{three} states that evolve cyclically, and calculate the (phase invariant) BP as in equation \ref{BP}.  In terms of amplitudes, we have
\begin{equation}
\gamma = -\textrm{Im} \ln \phi_{13} \; \phi_{32} \; \phi_{21}.
\end{equation}
Because the BP is independent of phase choice, so too is the product of amplitudes $\phi_{13} \; \phi_{32} \; \phi_{21}$.  This suggests the definition
\begin{equation}
\Gamma \equiv \phi_{1N} \cdot \cdot \cdot \phi_{32} \phi_{21} = \displaystyle\prod_{n=1}^N \phi_{i+1, i},
\end{equation}
where it is understood that the loop eventually closes upon itself so that $N+1 \rightarrow 1$.  So defined, $\Gamma$ is not a phase, but (like the BP) it is phase invariant; moreover, $\Gamma$ (what we call a ``product of amplitudes'') will appear naturally in quantum systems.

To see how this is so, we make the following observation: \emph{the probability of going from an initial state $\ket{i}$ to a final state $\ket{f}$ is the sum of all the $\Gamma$'s that include $\ket{i}$ and $\ket{f}$}.  Explicitly,
\begin{equation}
P_{fi} = \displaystyle\sum_{\mathcal{S}} \Gamma(\mathcal{S}),
\end{equation}
where the $\mathcal{S}$'s represent all possible closed paths that include $\ket{i}$ and $\ket{f}$.
For example, suppose a state $\ket{i}$ goes to a state $\ket{f}$ via only one possible (intermediate) route $\ket{1}$:
\begin{equation}
\ket{i} \rightarrow \ket{1} \rightarrow \ket{f}.
\end{equation}
Traditionally we would write, for the amplitude of an $\ket{i} \rightarrow \ket{f}$ transition via $\ket{1}$,
\begin{equation}
\langle f \ket{i} = \langle f \ket{1} \langle 1 \ket{i} = \phi_{f1} \; \phi_{1i} \equiv \phi_{fi,1},
\end{equation}
so that the probability of such a transition is simply $P_{fi}=|\phi_{fi,1}|^2$.  By our ``$\Gamma$-rule'', we get the same result: there is only one possible closed loop,
\begin{equation}
\ket{i} \rightarrow \ket{1} \rightarrow \ket{f} \rightarrow \ket{1} \rightarrow \ket{i},
\end{equation}
and we get for the total probability
\begin{equation}
P_{fi}= \langle i \ket{1} \langle 1 \ket{f} \langle f \ket{1} \langle 1 \ket{i} = \phi_{fi,1}^* \phi_{fi,1} = |\phi_{fi,1}|^2,
\end{equation}
as before.  (Note that there are not yet any interference terms as in equation \ref{interference}.)  We resist the temptation at this point to interpret the meaning of such a closed loop; for now we consider it a convenient mathematical device.

To get interference, we require there to be \emph{two} means to go from $\ket{i}$ to  $\ket{f}$: via $\ket{1}$ or $\ket{2}$.  Traditionally, one finds that
\begin{equation}
\langle f \ket{i} = \langle f \ket{1} \langle 1 \ket{i} + \langle f \ket{2} \langle 2 \ket{i},
\end{equation}
giving a probability of
\begin{equation} \label{probability}
P_{fi}=|\phi_{fi,1}|^2 + |\phi_{fi,1}|^2 + \phi_{fi,1} \phi_{fi,2}^* + \phi_{fi,1}^* \phi_{fi,2}
\end{equation}
for the $\ket{i}$ to  $\ket{f}$ transition.

Observe that interference terms are now present: these are the ``mixed'' terms $\phi_{fi,1} \phi_{fi,2}^*$, and $\phi_{fi,1}^* \phi_{fi,2}$.  These are the terms that supposedly have no ``classical'' interpretation.  Indeed, if we insist on reading $P_{fi}$ as a ``probability of a state $i$ evolving into state $f$'' then the interference terms remain classically inexplicable.  After all, routes $1$ and $2$ are mutually exclusive, so there is no obvious reason why $P_{fi}$ should depend in any way on such ``mixed'' terms.  Since equation \ref{classical law} does not hold, we are left with interference as the epitome of ``quantum weirdness'' \cite{Knight}.

Our ``$\Gamma$-rule'', however, provides an equivalent way to calculate $P_{fi}$.  We have four possible closed loops,
\begin{eqnarray}
\ket{i} \rightarrow \ket{1} \rightarrow \ket{f} \rightarrow \ket{1} \rightarrow \ket{i}, \nonumber \\
\ket{i} \rightarrow \ket{2} \rightarrow \ket{f} \rightarrow \ket{2} \rightarrow \ket{i}, \nonumber \\
\ket{i} \rightarrow \ket{1} \rightarrow \ket{f} \rightarrow \ket{2} \rightarrow \ket{i}, \nonumber \\
\ket{i} \rightarrow \ket{2} \rightarrow \ket{f} \rightarrow \ket{1} \rightarrow \ket{i},
\end{eqnarray}
with the respective probabilities $|\phi_{fi,1}|^2$, $|\phi_{fi,1}|^2$, $\phi_{fi,1} \phi_{fi,2}^*$, and $\phi_{fi,1}^* \phi_{fi,2}$.  Thus the total probability $P_{fi}$ is found to be the same as in equation \ref{probability}.  Perhaps this is mathematical coincidence, but the presence of the phase invariant BP-like quantity $\Gamma$ suggests another explanation.  It is natural to conclude from this analysis that closed loops are in some sense \emph{more fundamental} than transitions between states.  The authors of \cite{Anastopoulos} hint at this, but shy away from our hypothesis.  Going one step further, we suggest the following: that \emph{in deciding on the probability of a transition $\ket{i}$ to  $\ket{f}$, one can simply choose equally (and classically) from all possible closed loops that contain $\ket{i}$ and $\ket{f}$}. In other words, $P_{fi}$ can be considered to be the (classical) sum of \emph{four} mutually exclusive alternatives, as opposed to being a sum of two alternatives with two interference terms tacked on.  The question remains whether these alternatives have any meaningful philosophical interpretation.

\section{A classical analogy}
\label{sec:Analogy}
To get the flavor of this idea, consider the following parable, which is meant to act as an ``intuition pump'' \cite{Dennett} (and is not meant to be taken as a literal representation of QI).  A man lives in a house out of town, but visits town once a day.  There is only one viable road ($A$) between his house and the town, so his round trip (a closed loop) is always the same.  If we use an asterisk $*$ to denote the road used on his return, then there is only one possible round trip to consider, $AA^*$.

Now suppose that a different road ($B$) is built that gives the man another way to get to town.  In the traditional language of (quantum) transitions, we would say that the man's options have doubled, since there are two routes ($A$ \& $B$) where before there was one.  Within the conceptual framework of our product of amplitudes ($\Gamma$), however, we see that there are now \emph{four} possible closed loops: $AA^*$, $BB^*$, $AB^*$, and $BA^*$.  Let us suppose that the man chooses among these four possibilities equally before he sets off on his daily journey.  The marginal probability of $AA^*$ is now different, since $P(AA^*)=\frac{1}{4}$ instead of 1, but there is still no ``interference''; the transition ``HOME $\rightarrow$ TOWN via $A$'' occurs with probability $\frac{1}{2}$ under either interpretation.

Things become more complicated if we split the $A$ road into subroads $1$ and $2$, and likewise split $B$ into subroads $2$ and $3$.
\begin{figure*}
\includegraphics[width=0.75\textwidth]{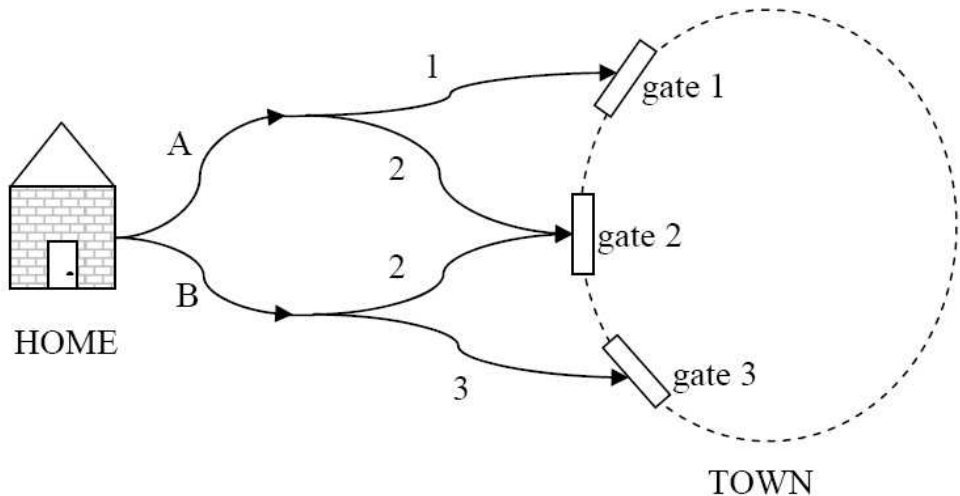}
\caption{Schematic representation of a parable which can exhibit interference.}
\end{figure*}
In this notation the numbers refer to any of three gates through which the roads can enter town (gate $2$ being common to both roads).  We will require that the man always leave town through the same gate that he entered; this reduces the number of possible round trips to six ($A1A^*$, $A2A^*$, $A2B^*$, $B2A^*$, $B2B^*$, $B3B^*$) and makes tabulations more manageable without loss of generality.

Now at each gate is a box filled with coins; visitors are supposed to pay upon entry into town.  If our visitor is diligent (meaning that he always pays a coin upon entry) then the number of coins in gate box $2$ will increase more rapidly (with each passing day) than the number of coins in gate boxes $1$ or $3$.  Indeed, a consideration of round trips indicates that gate $2$ is visited with a probability of $\frac{2}{3}$.

The day-to-day behavior of this man would be mysterious to an outside observer steeped in quantum orthodoxy.  Such an observer would notice that gate box $2$ gains coins at four times the rate of either of the other two gate boxes.  This contradicts the ``quantum transition'' viewpoint, in which one would expect to see coins increase as binomial coefficients, i.e. $1$ in gate box $1$, $2$ in gate box $3$, and $1$ in gate box $3$.  Additionally, if road $B$ were subsequently closed, an observer would see this unexpected behavior disappear, since gate boxes $1$ and $2$ would then gain coins at the same rate (such as in our less complicated initial example).

If, however, our observer steps back and considers round trips (closed loops), then things are more clear.  With both roads open, the observer would note that the man is simply choosing among six possible round trip alternatives.  With each alternative equally likely, gate $2$ is visited two thirds of the time.  With one road closed, the number of round trips is only four, and gates $1$ and $2$ are equally likely.  The observer concludes there is nothing mysterious going on; the situation is understandable if one examines round trips.

To mimic actual destructive interference, we introduce a rule that lets the traveler distinguish between different types of round trips.  Suppose our visitor is fickle instead of diligent; he sometimes removes a coin at the gate box instead of paying.  We stipulate the following rules for the fickle observer:
\begin{enumerate}
\item If the man intends to return in exactly the \emph{same} way as he came, he \emph{adds} a coin to the gate box.
\item If the man intends to return in a \emph{different} way than he came, he \emph{removes} a coin.
\end{enumerate}
Under these rules, the man will always pay a coin if he arrives via gate $1$ or gate $3$, since he has no choice but to return by the same way that he came.  If he arrives at gate $2$, however, he will take a coin (for the ``exotic'' routes $A2B^*$ and $B2A^*$) as often as he pays (for the ``boring'' routes $A2A^*$ and $B2B^*$).  On average, therefore, the number of coins in gate box $2$ will stay the same.  To an orthodox observer, this looks like destructive interference.  Such interference only appears when we make a distinction between exotic round trips and boring ones: the journey \emph{to town} must be different from the journey \emph{back home}.
\begin{figure} \label{coins}
Diligent:
\begin{tabular}{|c|c|}
\hline
Gate & Coins \\
\hline
$1$ & $1/6$ \\
$2$ & $2/3$ \\
$3$ & $1/6$ \\
\hline
\end{tabular}
$\; \; \; \;$ Fickle:
\begin{tabular}{|c|c|}
\hline
Gate & Coins \\
\hline
$1$ & $1/2$ \\
$2$ & 0 \\
$3$ & $1/2$ \\
\hline
\end{tabular}
\caption{Average number of coins given to each gate per visit, for a diligent or fickle town visitor.  Neither case is explicable without examining round trips (closed loops).}
\end{figure}

An outside observer would find the fickle man's behavior inexplicable because of this apparent destructive ``interference'', a decidedly non-classical phenomenon.  But a closed loop viewpoint leads an observer to decide that there really isn't any interference after all; the traveler is choosing between round trips, and changing his behavior based upon whether those round trips are boring or exotic.  In both the diligent and fickle cases it is the closed loop perspective that makes the data explicable in classical terms.

While admittedly artificial, this system behaves analogously to the toy quantum system considered before.  What appears to be ``interference'', unexplainable in terms of a transition HOME $\rightarrow$ TOWN, becomes explainable and classical if one looks at round trips instead.

\section{Quantum decoherence}
\label{sec:Decoherence}
Interestingly, a closed loop viewpoint can also provide insight into the phenomenon of quantum decoherence \cite{Zurek}.  By quantum decoherence we mean the process by which ``mixed'' probability terms (such as $\phi_{fi,1} \phi_{fi,2}^*$, and $\phi_{fi,1}^* \phi_{fi,2}$) vanish when a system interacts with its environment.  This is not full-fledged wave function collapse \`{a} la the Copenhagen interpretation; decoherence (and its consequence, einselection) produces an \emph{effective} wave function collapse \cite{Zurek}.  Essentially, the environmental interaction (a ``measurement'') rules out certain ``mixed'' amplitudes, and we (macroscopically) see the disappearance of QI.

Consider again the quantum system that evolves from $\ket{i}$ to $\ket{f}$ via one of the intermediate states $\ket{1}$ or $\ket{2}$.  There are two distinct possibilities for state $\ket{f}$: either
\begin{enumerate}
\item $\ket{f}$ does not depend \emph{in any way} on the intermediate state by which the system ``arrived'' at $\ket{f}$, or
\item $\ket{f}$ \emph{does} depend upon the intermediate state.
\end{enumerate}
In the first case, all four possible closed loops (denoted by $11^*$, $12^*$, $21^*$, and $22^*$) are consistent with time reversal.  There is no contradiction in supposing (for example) that a system could evolve as $\ket{i} \rightarrow \ket{1} \rightarrow \ket{f} \rightarrow \ket{2} \rightarrow \ket{i}$, since the state $\ket{f}$ contains no information about the path taken.  In the second case, however, the ``mixed'' routes $12^*$ and $21^*$ lead to time reversal ``paradoxes''.  A state $\ket{f}$ that has (as one of its properties) \emph{information about how the state was achieved}, cannot then evolve back into state $\ket{i}$ via a different path.  The only choices then available are the closed loops $11^*$ and $22^*$; as a consequence the QI probability terms have vanished.  In effect, the making of any measurement rules out any ``mixed'' closed loops, which do not then enter into any probability calculations.

This is not to say that a measurement (and consequent decoherence) actively \emph{changes} a state that does \emph{not} depend on intermediate states to a different state that does.  Rather, from a many-worlds viewpoint \cite{Everett},\cite{Tegmark} a measurement rules out the possibility that we are in a ``mixed'' state universe.  For the above system, if all we know is that a transition $\ket{i} \rightarrow \ket{f}$ takes place, then there is no way in general to choose among the four possibilities $11^*$, $12^*$, $21^*$, and $22^*$; each is equally likely to represent the universe we are in.  But if the state $\ket{f}$ depends in some way upon the path taken, then (upon finding a state $\ket{f}$) we must be in either the ``$11^*$'' universe or the ``$22^*$'' universe.  Decoherence has been achieved by a classical selection from available closed loops.  In the language of our parable, the state $\ket{f}$ must be executing a boring round trip since an exotic round trip would lead to paradox.

\section{Possible interpretations}
\label{sec:Possible}
We now make a few remarks regarding this discussion.  Firstly, all quantum transitions can be couched in the language of our product of amplitudes ($\Gamma$), and this quantity (like the Berry's phase) is always phase invariant.  Therefore, regardless of whether or not we believe that the systems in question are actually traversing closed loops in time, we have yet another mathematical framework in which to interpret quantum probabilities.  But it is precisely the invariant nature of $\Gamma$ that suggests we go further: it is only through closed loops in time that QI can be interpreted classically, so perhaps such loops are important.  In the language of the Berry's phase, perhaps time itself should be considered a parameter of a state's governing Hamiltonian, and probabilities should be calculated based upon the assumption that all states eventually ``loop back on themselves''.

Of course, treating time this way is problematic, for it is not obvious what a closed loop in time \emph{means}.  (Indeed, such a supercyclic time requires us to abandon principles of causality: future events can be the ``cause'' of something in the past \cite{Raju}.)  But for visualization purposes, there is no mathematical difference between a Feynman-like diagram in which states loop back onto themselves and a linear diagram that shows states being periodic.  Both evolutions yield identical BP's and identical $\Gamma$'s.
\begin{figure*}
\includegraphics[width=0.75\textwidth]{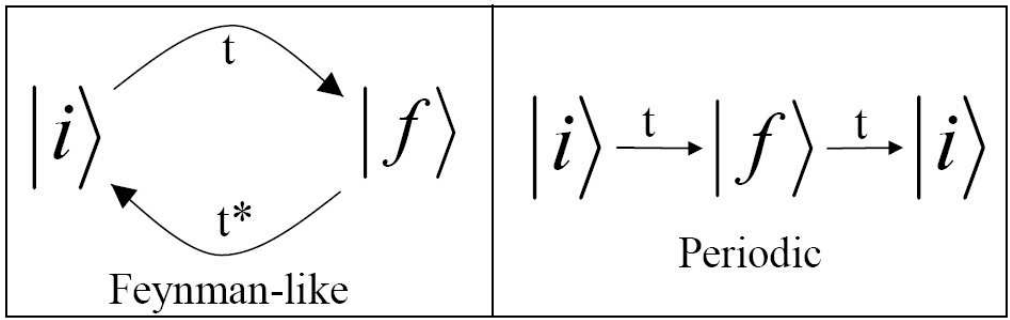}
\caption{Two ways to visualize a closed loop in time.  The first way , with a forward time $t$ and a ``backward time'' $t^*$, seems nonsensical.  The second way is mathematically equivalent.}
\end{figure*}
In light of this, we speculate that the appearance of closed timelike curves \cite{Godel}, \cite{Gott} in certain solutions to Einstein's field equation (such as in the Kerr metric \cite{Kerr}) are not mere artifacts.  Rather, they might be provide evidence that the closed loop viewpoint of time is not merely a mathematical convenience, but may have some independent significance.  This ties in with recent speculation that the universe itself may be time symmetric on large (cosmological) scales \cite{Carroll}.

Secondly, one may ask how this interpretation differs from Feynman's path integral formulation \cite{Feynman2} of quantum mechanics.  The answer is that it differs only in viewpoint.  According to the path integral prescription, the probability amplitude of a particular transition is calculated by adding up all of the individual amplitudes (histories) that include the initial and final states.  There is no ``reason'' for this in terms of classical probability, and to find the probability one must still take the squared modulus of a (presumably) complex amplitude.  In light of the phase invariant product of amplitudes, however, we see that there is another viewpoint.  We consider as fundamental not only histories, but futures, and include all such history/future combinations.  When doing this, ``mixed'' terms (ostensibly due to interference) are seen to be (classical) mutually exclusive alternatives.  If taken at face value, this suggests that all states are periodic, in the sense that all history/future paths are closed.  With these assumptions, classical probability is thereby restored, at the expense of introducing periodicity to quantum systems.  Whether this is just a mathematical trick, or instead has real philosophical implications, remains an area for future investigation.

\section{Conclusions}
\label{sec:Conclusions}
We have shown that a Berry's phase-like quantity, $\Gamma$, can be defined for any quantum system which undergoes a transition.  This product of amplitudes is phase invariant, in the sense that it does not matter what phase choices are made for individual kets.  We have also shown how all quantum probabilities can be classically interpreted in terms of a sum of $\Gamma$'s, regardless of whether or not we believe that the systems in question are \emph{actually} traversing closed loops in time.  If we accept such loops, then quantum interference can be classically interpreted, and the ``mystery'' of quantum probabilities is replaced with the ``mystery'' of how time can loop back on itself.  If we are uncomfortable with such loops, then the product of amplitudes will remain a mathematical curiosity, and the ``mystery'' of quantum interference will remain a subject of debate.

\begin{acknowledgements}
I wish to thank William Hodge, Scott Huffman, and William Kerr for helpful discussions.
\end{acknowledgements}

%Hamburger, C.: Quasimonotonicity, regularity and duality for nonlinear systems of partial differential equations. Ann. Mat. Pura. Appl. 169, 321–354 (1995)
%Geddes, K.O., Czapor, S.R., Labahn, G.: Algorithms for Computer Algebra. Kluwer, Boston (1992)
%Seymour, R.S. (ed.): Conductive Polymers. Plenum, New York (1981)
%Broy, M.: Software engineering – from auxiliary to key technologies. In: Broy, M., Denert, E. (eds.) Software Pioneers, pp. 10–13. Springer, Berlin Heidelberg New York %(2002)

\end{document}